\documentclass[11pt]{article}

\usepackage[cp1251]{inputenc}

\newcommand{\N}{N\raise.7ex\hbox{\underline{$\circ $}}$\;$}

\begin{document}

\title{
E.M. Ovsiyuk\footnote{e.ovsiyuk@mail.ru}  \\
On finding parameters of Mueller matrices of the Lorentzian type  from results
of polarization measurements \\
{\small Mozyr State Pedagogical University named after I.P. Shamyakin, Belarus
}}

\maketitle

\begin{abstract}

With assumption that an optical element is described by a Mueller
matrix of the Lorentzian type, a method to find a 3-dimensional
complex vector-parameter  for
 a corresponding Mueller matrix from results of
  four specially chosen polarization measurements has been elaborated.

\end{abstract}

 It is known that the four Stokes parameters describe the
state of polari\-zation of the  light. They were introduced by Stokes in
1852 \cite{Stokes-1852}. Mueller calculus is a matrix method for  dealing with
Stokes 4-vectors,
 it was developed in 1943 by H. Mueller \cite{Mueller-1943}. Any optical element can be represented
 by a Mueller matrix. In optics, polarized light can be described using the Jones calculus,
invented by Jones in 1941 \cite{Jones-1941}--\cite{Jones-1947}.
Polarized light is represented by a 2-dimensional Jones complex vector,
and linear optical elements are represented by $2 \times 2$ Jones matrices.
  The Jones calculus is only applicable to light that is  completely  polarized.
  Commonly, light which is partially polarized is treated only  with the use of vector Mueller calculus.

It is  well known    that when describing (completely or partially)   polarized  light
a noticeable role may  be given to the group of $(3+1)$-pseudo orthogonal transformations
consisting of a group $SO(3,1) $ (isomorphic to the Lorentz group). Therefore,
techniques developed  in the frames of the Lorentz group (for instance, see
 \cite{Gel'fand-1963}--\cite{Bogush-Red'kov-2008}), in particular within relativistic kinematics, may play heuristic
  role in exploring  optical problems (the bibliography  on the subject is enormous, many
references are given in \cite{Bogush-2007}--\cite{Red'kov-2011}).

In a previous paper \cite{Red'kov-2012},
  a general group-theoretic method for recovery of the Mueller matrices of the Lorentzian type
   for any  (Lorentzian type) optical element from  results of several independent polarization experiments
    was investigated. A main feature of treatment given in  \cite{Red'kov-2012} is
     that  initial (probing) beams of light are  arbitrary.
     Meanwhile, in the book by Snopko \cite{Snopko-1992},  an entirely different way to restore a 16-element Mueller matrices
      (not necessarily of the Lorentzian type) is described.
This  method is based on the use of specially chosen probing beams of light:
one natural and three completely  polarized.

A natural question about the correlation of these two
 techniques\footnote{The author is grateful to E.A. Tolkachev and Y.A. Kurochkin for pointing out on this fact.
} arises. The main goal of  the present paper is to  investigate restriction of  the general method [18]
to the special case of Mueller matrices of the Lorentzian type.

Let us describe the general rules  for finding Mueller matrices of an arbitrary optical element
 (first without any restriction to the class of  Mueller matrices
 of the Lorentzian type) [18]. The first probing light  beam  is chosen as the  natural light
$$
S^{a}_{(0)} = (I, 0, 0, 0) \qquad  \Longrightarrow \qquad
S^{a'}_{(0)}\; ,
$$
$$
M = \left | \begin{array}{cccc}
m_{00} & m_{01} & m_{02} & m_{03} \\
m_{10} & m_{11} & m_{12} & m_{13} \\
m_{20} & m_{21} & m_{22} & m_{23} \\
m_{30} & m_{31} & m_{32} & m_{33}
\end{array} \right |
\left | \begin{array}{c}
I = S^{0}_{(0)}\\
0   \\
0  \\
0
\end{array} \right |  =
\left | \begin{array}{c}
 S^{0'}_{(0)}\\
 S^{1'}_{(0)}\\
 S^{2'}_{(0)}
\\
 S^{3'}_{(0)}
\end{array} \right | ,
$$
$$
m_{00} I    =  S^{0'}_{(0)} \;, \qquad m_{10}  I =  S^{1'}_{(0)}
\; , \qquad m_{20} I =  S^{2'}_{(0)} \; , \qquad m_{30} I =
S^{3'}_{(0)}  \; . \eqno(1)
$$

The next three probing beams are chosen as completely polarized ones, and of a special form:
$$
S^{a}_{(1)} = (I, I, 0, 0)\qquad  \Longrightarrow \qquad
S^{a'}_{(1)}\; ,
$$
$$
M = \left | \begin{array}{cccc}
m_{00} & m_{01} & m_{02} & m_{03} \\
m_{10} & m_{11} & m_{12} & m_{13} \\
m_{20} & m_{21} & m_{22} & m_{23} \\
m_{30} & m_{31} & m_{32} & m_{33}
\end{array} \right |
\left | \begin{array}{c}
I \\
I     \\
0  \\
0
\end{array} \right |  =
\left | \begin{array}{c}
 S^{0'}_{(1)}\\
S^{1'}_{(1)}\\
 S^{2'}_{(1)}
\\
 S^{3'}_{(1)}
\end{array} \right |,
$$
$$
S^{0'}_{(0)} + m_{01} I = S^{0'}_{(1)}\; , \qquad S^{1'}_{(0)}  +
m_{11} I = S^{1'}_{(1)} \; ,
$$ $$
S^{2'}_{(0)}  + m_{21} I = S^{2'}_{(1)} \; , \qquad S^{3'}_{(0)}
+ m_{31} I = S^{3'}_{(1)} \; ; \eqno(2)
$$

$$
S^{a}_{(2)} = (I, 0, I, 0)\qquad  \Longrightarrow \qquad
S^{a'}_{(2)}\; ,
$$
$$
M = \left | \begin{array}{cccc}
m_{00} & m_{01} & m_{02} & m_{03} \\
m_{10} & m_{11} & m_{12} & m_{13} \\
m_{20} & m_{21} & m_{22} & m_{23} \\
m_{30} & m_{31} & m_{32} & m_{33}
\end{array} \right |
\left | \begin{array}{c}
I \\
0     \\
I  \\
0
\end{array} \right |  =
\left | \begin{array}{c}
 S^{0'}_{(2)}\\
S^{1'}_{(2)}\\
 S^{2'}_{(2)}
\\
 S^{3'}_{(2)}
\end{array} \right |,
$$
$$
S^{0'}_{(0)} + m_{02} I = S^{0'}_{(2)}\; , \qquad S^{1'}_{(0)}  +
m_{12} I = S^{1'}_{(2)} \; ,
$$ $$
S^{2'}_{(0)}  + m_{22} I = S^{2'}_{(2)} \; , \qquad S^{3'}_{(0)}
+ m_{32} I = S^{3'}_{(2)} \; ; \eqno(3)
$$

$$
S^{a}_{(3)} = (I, 0, 0, I)\qquad  \Longrightarrow \qquad
S^{a'}_{(3)}\; ,
$$
$$
M = \left | \begin{array}{cccc}
m_{00} & m_{01} & m_{02} & m_{03} \\
m_{10} & m_{11} & m_{12} & m_{13} \\
m_{20} & m_{21} & m_{22} & m_{23} \\
m_{30} & m_{31} & m_{32} & m_{33}
\end{array} \right |
\left | \begin{array}{c}
I \\
0     \\
0  \\
I
\end{array} \right |  =
\left | \begin{array}{c}
 S^{0'}_{(3)}\\
S^{1'}_{(3)}\\
 S^{2'}_{(3)}
\\
 S^{3'}_{(3)}
\end{array} \right |,
$$
$$
S^{0'}_{(0)} + m_{03} I = S^{0'}_{(3)}\; , \qquad S^{1'}_{(0)}  +
m_{13} I = S^{1'}_{(3)} \; ,
$$ $$
S^{2'}_{(0)}  + m_{23} I = S^{2'}_{(3)} \; , \qquad S^{3'}_{(0)}
+ m_{33} I = S^{3'}_{(3)} \; . \eqno(4)
$$

The resulting system of equations leads to the following explicit expressions for the
16 elements of the Mueller matrix
$$
m_{00}     =  S^{0'}_{(0)} / I \;, \; m_{10}   =  S^{1'}_{(0)}
/ I \; , \; m_{20}  =  S^{2'}_{(0)} / I \; , \; m_{30}  =
S^{3'}_{(0)} / I  \; ,
$$
$$
  m_{01}  = {S^{0'}_{(1)} - S^{0'}_{(0)} \over  I}  , \;
 m_{11}  = {S^{1'}_{(1)} - S^{1'}_{(0)} \over I}  ,\;
 m_{21}  = {S^{2'}_{(1)} - S^{2'}_{(0)} \over I} , \;
 m_{31}  = {S^{3'}_{(1)} - S^{3'}_{(0)} \over  I}  ,
$$
$$
 m_{02}  = { S^{0'}_{(2)} - S^{0'}_{(0)} \over I}  , \;
 m_{12}  = {S^{1'}_{(2)}  - S^{1'}_{(0)} \over I  } ,
\; m_{22}  = {S^{2'}_{(2)} - S^{2'}_{(0)} \over I}  , \;
 m_{32}  = {S^{3'}_{(2)} - S^{3'}_{(0)} \over I}  ,
$$
$$
m_{03}  = {S^{0'}_{(3)}- S^{0'}_{(0)} \over I}  , \;
 m_{13}  = {S^{1'}_{(3)} - S^{1'}_{(0)}  \over I}   ,
\;
 m_{23}  = {S^{2'}_{(3)}  - S^{2'}_{(0)} \over I}  , \;
 m_{33}  = {S^{3'}_{(3)} - S^{3'}_{(0)}  \over I}  .
$$
$$
\eqno(5)
$$

We consider the case when the matrix is isomorphic to the Mueller matrix
 of the Lorentz group.
 This means that we must assume that  all  Stokes
 vectors are invariant  in a sense of  "relativistic length": $ g_ {ab} S ^ {a} S ^ {b} = g_ {ab} S ^ {a '} S ^ {b' } $, so we get
$$
I^{2} =   (S^{0'}_{(0)})^{2} - ( {\bf S}'_{(0)} )^{2} \; , \qquad
0 =  (S^{0'}_{(1)})^{2} - ( {\bf S}'_{(1)} )^{2} \; ,
$$
$$
0 =  (S^{0'}_{(2)})^{2} - ( {\bf S}'_{(2)} )^{2} \; , \qquad
0 =  (S^{0'}_{(3)})^{2} - ( {\bf S}'_{(3)} )^{2} \; . \eqno(6)
$$

We first consider a simpler problem,
assuming that the matrix is isomorphic to the element of a group of 3-dimensional rotations:
$$
M = \left | \begin{array}{cccc}
1 & 0 & 0 & 0  \\
0 & m_{11} & m_{12} & m_{13} \\
0 & m_{21} & m_{22} & m_{23} \\
0  & m_{31} & m_{32} & m_{33}
\end{array} \right |.
\eqno(7)
$$

\noindent The system (5) takes the form
$$
1      =  1  \;, \qquad 0   =  0  \; , \qquad 0  =  0 \; , \qquad
0  =  0  \; ,
$$
$$
 0  = 0\; , \qquad
 m_{11}  = {S^{1'}_{(1)}  \over I} \; ,\qquad
 m_{21}  = {S^{2'}_{(1)}  \over I}\; , \qquad
 m_{31}  = {S^{3'}_{(1)}  \over  I} \; ,
$$
$$
 0  = 0\; , \qquad
 m_{12}  = {S^{1'}_{(2)}   \over I  }\; ,
\qquad m_{22}  = {S^{2'}_{(2)}  \over I} \; , \qquad
 m_{32}  = {S^{3'}_{(2)}  \over I} \; ,
$$
$$
0= 0 \; , \qquad
 m_{13}  = {S^{1'}_{(3)}   \over I}  \; ,
\qquad
 m_{23}  = {S^{2'}_{(3)}  \over I} \; , \qquad
 m_{33}  = {S^{3'}_{(3)}  \over I} \; .
\eqno(8)
$$

\noindent Additional conditions (6) are  simplified
$$
I^{2} =   I^{2}  \; , \qquad
I^{2} =  ( {\bf S}'_{(1)} )^{2} \; , \qquad
I^{2} =  ( {\bf S}'_{(2)} )^{2} \; , \qquad
I^{2} =  ( {\bf S}'_{(3)} )^{2} \; . \eqno(9)
$$

According to (8), the matrix (7) can be represented in the form (we follow only the three-dimensional matrix)
$$
M = {1 \over I } \left | \begin{array}{ccc}
 S^{1'}_{(1)}  & S^{1'}_{(2)}  & S^{1'}_{(3)} \\
 S^{2'}_{(1)} & S^{2'}_{(2)} &  S^{2'}_{(3)} \\
 S^{3'}_{(1)}  & S^{3'}_{(2)} & S^{3'}_{(3)}
\end{array} \right |  .
\eqno(10)
$$

\noindent The determinant of this matrix (belonging  to rotation group) must be equal to $ 1 $
$$
\mbox{det}\; M = {1 \over I } \left | \begin{array}{ccc}
 a_{1}   &  b_{1}   &  c_{1}  \\
 a_{2}   &  b_{2}   &  c_{2}  \\
 a_{3}   &  b_{3}   &  c_{3}
\end{array} \right |  = {1 \over I^{3} } \; {\bf a} ( {\bf b} \times {\bf c})  = 1 \; .
\eqno(11a)
$$

\noindent
If we use the polarization vector $ {\bf S} = I {\bf p} $, then the resulting constraint can be expressed as
$$
{\bf p}'_{(1)} ( {\bf p}'_{(2)} \times {\bf p}'_{(3)})  = 1\; .
\eqno(11b)
$$

\noindent
These three vectors cannot be considered as independent quantities
 since they are  obtained as a result of rotating of three  initial polarization vectors:
$$
{\bf p}_{(1)} = (1,0,0)\;, \qquad {\bf p}_{(2)} = (0,1,0)\;,
\qquad {\bf p}_{(3)} = (0,0,1)\; . \eqno(11c)
$$

\noindent
Three vectors $ {\bf p }'_{( 1)}, {\bf p }'_{( 2)}, {\bf p }'_{( 3)} $
must have magnitude 1, orthogonal to each other, and be  right-handed ones.

Matrix (10) must be identified with the orthogonal matrix of the group $SO (3, R)$ (see in [10])
$$
O = \left | \begin{array}{lll}
 1 -2 (n_{2}^{2} + n_{3}^{2})  & \qquad
 -2n_{0}n_{3} + 2n_{1}n_{2}         & \qquad
 +2n_{0}n_{2} + 2n_{1}n_{3}         \\[2mm]
 +2n_{0}n_{3} + 2n_{1}n_{2}        & \qquad
 1 -2 (n_{3}^{2} + n_{1}^{2})  & \qquad
 -2n_{0}n_{1} + 2n_{2}n_{3}         \\[2mm]
 -2n_{0}n_{2} + 2n_{1}n_{3}         &\qquad
+2n_{0}n_{1} + 2n_{2}n_{3}         &\qquad
 1 -2 (n_{1}^{2} + n_{2}^{2})  ,
\end{array} \right |,
\eqno(12a)
$$

\noindent parameters satisfy the condition
$$
n_{0}^{2} + n_{1}^{2} + n_{2}^{2} + n_{3}^{2} = +1  \; .
\eqno(12b)
$$

Further,  we use (with minor modifications) technique developed in \cite{Fedorov-1979}.
 We compute $ \mbox {Sp} \; M $ and find $ n_ {0} $:
$$
\mbox{Sp}\; M = (S^{1'}_{(1)} + S^{2'}_{(2)}+ S^{3'}_{(3)})/I=
p^{1'}_{(1)} + p^{2'}_{(2)}+ p^{3'}_{(3)} \; ,
$$
$$
2n_{0} = \sqrt{ p^{1'}_{(1)} + p^{2'}_{(2)}+ p^{3'}_{(3)} +1}   \;
. \eqno(13)
$$

\noindent Let us separate  an  antisymmetric part of the matrix $ M_ {as} = (M -
\tilde {M}) / 2 $ and equate it to the $ O_ {as} = (O - \tilde {O}) / 2 $
$$
{1 \over 2} \left | \begin{array}{ccc}
 0                             & -(p^{2'}_{(1)}- p^{1'}_{(2)}  )   & (p^{1'}_{(3)} -p^{3'}_{(1)} ) \\
 (p^{2'}_{(1)} -p^{1'}_{(2)})  &           0                       &  -(p^{3'}_{(2)} -p^{2'}_{(3)})  \\
 -(p^{1'}_{(3)} -p^{3'}_{(1)}) & (p^{3'}_{(2)} -p^{2'}_{(3)})      &              0
 \end{array} \right |
 $$
 $$
 =
 \left | \begin{array}{lll}
 0 & \qquad  -2n_{0}n_{3}          & \qquad  +2n_{0}n_{2}          \\[2mm]
 +2n_{0}n_{3}         & \qquad
 0 & \qquad
 -2n_{0}n_{1}        \\[2mm]
 -2n_{0}n_{2}        &\qquad
+2n_{0}n_{1}        &\qquad
 0
\end{array} \right |.
$$

\noindent
As a result, we obtain
$$
2n_{0}n_{1}  ={1 \over 2} (p^{3'}_{(2)} -p^{2'}_{(3)}) \; \;
\Longrightarrow \;\; n_{1}  ={ p^{3'}_{(2)} -p^{2'}_{(3)} \over
2\sqrt{ p^{1'}_{(1)} + p^{2'}_{(2)}+ p^{3'}_{(3)} +1}   }\,,
$$
$$
2n_{0}n_{2}  = {1 \over 2} (p^{1'}_{(3)} -p^{3'}_{(1)} ) \; \;
\Longrightarrow \;\; n_{2}  = { p^{1'}_{(3)} -p^{3'}_{(1)} \over
2\sqrt{ p^{1'}_{(1)} + p^{2'}_{(2)}+ p^{3'}_{(3)} +1}   } \; ,
$$
$$
2n_{0}n_{3}  = {1 \over 2} (p^{2'}_{(1)} -p^{1'}_{(2)}) \; \;
\Longrightarrow \;\; n_{3}  = { p^{2'}_{(1)} -p^{1'}_{(2)} \over
2\sqrt{ p^{1'}_{(1)} + p^{2'}_{(2)}+ p^{3'}_{(3)} +1}   } \; .
$$
$$
\eqno(14)
$$

\noindent It is easily   verified the equality (12b)
$$
 [ p^{1'}_{(1)} + p^{2'}_{(2)}+ p^{3'}_{(3)} +1  ] +
 { (p^{3'}_{(2)} -p^{2'}_{(3)})^{2}  \over   p^{1'}_{(1)} + p^{2'}_{(2)}+ p^{3'}_{(3)} +1   }
 $$
 $$
 + { (p^{1'}_{(3)} -p^{3'}_{(1)})^{2} \over   p^{1'}_{(1)} + p^{2'}_{(2)}+ p^{3'}_{(3)} +1   }+
  {(p^{2'}_{(1)} -p^{1'}_{(2)} )^{2} \over   p^{1'}_{(1)} + p^{2'}_{(2)}+ p^{3'}_{(3)} +1   } =4\,.
\eqno(15)
 $$

Note that, according to the procedure \cite{Red'kov-2012},
to restore the Mueller matrices in 3-dimensional case, it suffices to use
 only two pairs of vectors. Here we employ three pairs.

Let us go back to the 4-dimensional Mueller matrices
$$
M = {1 \over I} \left | \begin{array}{cccc}
S^{0'}_{(0)}  & S^{0'}_{(1)} - S^{0'}_{(0)} & S^{0'}_{(2)} - S^{0'}_{(0)}  &  S^{0'}_{(3)} - S^{0'}_{(0)} \\
S^{1'}_{(0)}  & S^{1'}_{(1)} - S^{1'}_{(0)} & S^{1'}_{(2)} - S^{1'}_{(0)}  &  S^{1'}_{(3)} - S^{1'}_{(0)} \\
S^{2'}_{(0)}  & S^{2'}_{(1)} - S^{2'}_{(0)} & S^{2'}_{(2)} - S^{2'}_{(0)}  &  S^{2'}_{(3)} - S^{2'}_{(0)}  \\
S^{3'}_{(0)}  & S^{3'}_{(1)} - S^{3'}_{(0)} & S^{3'}_{(2)} -
S^{3'}_{(0)}  &  S^{3'}_{(3)} - S^{3'}_{(0)}
\end{array} \right |
\eqno(16)
$$

\noindent
and introduce the notation
$$
S^{a'}_{(0)} = F^{a}\;, \qquad S^{a'}_{(1)} = A^{a}, \qquad
S^{a'}_{(2)} = B^{a},\qquad S^{a'}_{(3)} = C^{a}, \eqno(17)
$$

\noindent then
$$
M = {1 \over I} \left | \begin{array}{cccc}
F^{0}  & \;\; A^{0} - F^{0} & \;\; B^{0} - F^{0}  &  \;\;  C^{0} - F^{0} \\
F^{1}  & \;\; A^{1} - F^{1} & \;\; B^{1} - F^{1}  &  \;\;  C^{1} - F^{1} \\
F^{2}  & \;\; A^{2} - F^{2} & \;\; B^{2} - F^{2}  &  \;\;  C^{2} - F^{2}  \\
F^{3}  & \;\; A^{3} - F^{3} & \;\; B^{3} - F^{3}  &  \;\;  C^{3} -
F^{3}
\end{array} \right |.
\eqno(18)
$$

\noindent
We identify this matrix with the matrix of the Lorentz group
 and use a method of finding  parameters of the matrix Lorentz described in \cite{Red'kov-bookI}[20].
 In the main points, it coincides with well-developed  technique given in \cite{Fedorov-1979},
 the differences are related  with the transition to spinor covering $ SL (2.C) $ for the
 Lorentz group $ L_ {+}^{ \uparrow} $. Let us briefly describe
 this recipe. Any  orthochronous Lorentz transformation can be represented as follows:
$$
L^{\;\; a}_{b} (k,\; k^{*}) = \bar{\delta }^{c}_{b} \; \left ( \;
- \delta^{a}_{c} \; k^{n} \; k^{*}_{n} \;  + \;  k_{c} \; k^{a*}
\; + \; k^{*}_{c} \; k^{a}\; + \; i\; \epsilon ^{\;\;anm}_{c}\;
k_{n} \; k^{*}_{m} \; \right ) \;  , \eqno(19)
$$

\noindent where $ \bar {\delta} ^ {c} _ {b} $ -- special
(different from the usual) the Kronecker delta-symbol
$$
\bar{\delta }^{c}_{b} =  \left \{ \begin{array}{l}
 0 , \; \; \;\; c  \neq  b \; ; \\
+1 , \; \;  \;\; c = b = 0 \; ; \\
-1 , \; \; \;\; c = b = 1, \; 2,\; 3 \; .
\end{array} \right.
$$

\noindent
We expand the four-dimensional parameter  $ k_ {a} $ into
 real and imaginary parts:
$$
k_{0}=  m_{0} \;  - \; i \; n_{0} = \Delta \; e^{i\kappa } \;
,\qquad {\bf k} = (k_{j}) = \; {\bf m} \; - \;  i\;  {\bf n} \;
\eqno(20)
$$

\noindent and represent the matrix $ \Lambda \; \; (L ^ {\; \; b} _ {a} =
\bar {\delta} ^ {c} _ {a} \; \Lambda ^ {\; \; b} _ {c}) $ as the  sum
of symmetric and antisymmetric parts of the $ \Lambda = (S \; + \; A) $:
$$
S = \left | \begin{array}{cc} \Delta ^{2}\; + \; {\bf m}^{2} \;+\;
{\bf n}^{2}  &
2 \; [ {\bf n} \; {\bf m}\; ]   \\
2\; [{\bf n}\; {\bf m} \;]      &
 -\Delta ^{2}\;+\;{\bf m}^{2}\; + \; {\bf n}^{2} \;-\; 2 \;
{\bf m}\;\bullet \; {\bf m} \; - \; 2 \;  {\bf n} \; \bullet\;
{\bf n}
\end{array} \right |  ,
$$
$$
A = 2 \; \Delta \; \left | \begin{array}{cl}
0      &  -( {\bf m} \; \cos  \kappa \; -\; {\bf n} \; \sin \kappa )  \\[2mm]
({\bf m} \; \cos  \kappa \;  - \;  {\bf n} \; \sin  \kappa ) &
\;\;\;({\bf m} \; \cos  \kappa \; + \; {\bf n} \; \sin  \kappa
)^{\times}
\end{array} \right |   .
\eqno(21)
$$

\noindent Notation is used: $ ({\bf n}\; \bullet \;
{\bf n})_{ij} = n_{i} \; n_{j} \; , \; ({\bf m} \;\bullet \;
{\bf m})_{ij}  = m_{i} \; m_{j}\; , \; ({\bf b}^{\times})_{ij} =
\epsilon _{ijk} \; b_{k}\; . $

Taking into account  the dependence of matrix elements of $ A $
on parameter $ \kappa $, the phase of a complex number $ k_ {0} $, we introduce
three-dimensional vectors $ {\bf M} $ and $ {\bf N} $
$$
\left | \begin{array}{c} {\bf M} \\   {\bf N}   \end{array} \right
|    = \left    | \begin{array}{cc}
\cos  \kappa    &   -  \sin  \kappa   \\
\sin  \kappa    &  \cos  \kappa  \end{array} \right | \; \left |
\begin{array}{c} {\bf m} \\ {\bf n} \end{array} \right | \; ,
\eqno(22)
$$

\noindent then  for the matrix $ S $ and $ A $ we obtain the representations
$$
S =  \left | \begin{array}{cc} \Delta ^{2} \; + \; {\bf M}^{2} \;
+ \; {\bf N}^{2} &
 2 \; [{\bf N}\; {\bf M} ] \\[2mm]
2\; [{\bf N} \; {\bf M}] &  -\Delta ^{2} \; + \; {\bf M}^{2}\; +
\; {\bf N}^{2} \;-\; 2 \; {\bf M}\;\bullet \;{\bf M} \;-\;
2\;{\bf N}\; \bullet \; {\bf N}  \end{array} \right |,
$$
$$
A = 2\; \Delta \;    \left | \begin{array}{cr}
0  &   - {\bf M}   \\[2mm]  +{\bf M}  &  \;\;{\bf N}^{\times}
\end{array} \right | \; .
\eqno(23)
$$

\noindent Relations $ (22) $ can be written in the form of a complex equation:
$$
e^{-i\kappa } \; {\bf k}  =  \; e^{-i\kappa } \; (\; {\bf m} \; -
\; i \;
 {\bf n} \;) =  {\bf M} \; - i\;  {\bf N}\; ,
$$

\noindent respectively, the constraint on  the determinant of the Lorentz matrix can be written as
$$
 \Delta ^{2} \; - \; (\;  {\bf M}  \;  -  \;  i   \; {\bf N} )^{2}
 =  e^{-2i\kappa }\; .
\eqno(24)
$$

Now, we formulate the rule for finding the explicit form of parameter $ k_ {a}$.
1)  First, since the equality holds
$$
\mbox{Sp}\; L = 2\; ( g^{nm} \;+\; \bar{g}^{nm}) \;
k_{n}\;k^{*}_{m} =
 4 \; k_{0}\;k^{*}_{0} = 4 \; \Delta ^{2} \; ,
\eqno(25)
$$

\noindent we must  compute  $ \mbox {Sp} \; L $ and then to find  the
value of $ \Delta$. 2) After that, by an  antisymmetric part $ A $  of the matrix
$ \Lambda $ we define the vectors ${\bf M}$  and  ${\bf N}$.
3) Finally, the found values so $(\; \Delta, \; {\bf M}, \;
{\bf N}\;)$ restore the parameter $k_{a}$
$$
( \; k_{0},\; k_{j}\; ) = \;
 { \pm 1 \over
\sqrt{\Delta^{2} \; - \; ( {\bf M} - i {\bf N} )^{2}} } \; \left (
\Delta  \; , \; {\bf M} \; - \; i\; {\bf N} \right ) \, ,
\eqno(26)
$$

\noindent where $ (\pm) $ represent the possibility of finding a spinor
transformation from the vector one only up to the sign $\pm$.

No we will  apply the formula  (26) to find the parameters of the matrix $ M $
(18). Compute the trace of the matrix $ M $ and the parameter $\Delta$:
$$
2\Delta =  \pm  { \sqrt{ F^{0}    + (A^{1} - F^{1})  + (B^{2}-
F^{2})  + ( C^{3}  - F^{3} ) \over I}}.
\eqno(27)
$$

\noindent
From the matrix $ M $ we obtain a matrix $\Lambda$
$$
\Lambda  = {1 \over I} \left | \begin{array}{rrrr}
F^{0}  & \;\; (A^{0} - F^{0}) & \;\; (B^{0} - F^{0})  &  \;\;  (C^{0} - F^{0}) \\
-F^{1}  & \;\; -(A^{1} - F^{1}) & \;\; -(B^{1} - F^{1})  &  \;\;-(  C^{1} - F^{1}) \\
-F^{2}  & \;\; -(A^{2} - F^{2}) & \;\; -(B^{2} - F^{2})  &  \;\;-(  C^{2} - F^{2} ) \\
-F^{3}  & \;\; -(A^{3} - F^{3} )& \;\; -(B^{3} - F^{3})  &  \;\;-(
C^{3} - F^{3})
\end{array} \right |.
$$

\noindent
Find the antisymmetric part of the matrix $\Lambda$ and identify it with
$$
  2\; \Delta \;    \left | \begin{array}{cccc}
0      &   - M_{1}  &  -M_{2}  &   - M_{3}   \\
M_{1}  &   0        &  N_{3}   &   -N_{2} \\
M_{2}  &   -N_{3}   &  0       &   N_{1}  \\
M_{2}  &   N_{2}    &  - N_{1} &   0
\end{array} \right | \; .
$$

\noindent
So we arrive at
$$
{ F^{0} - F^{1} - A^{0}   \over 2 I} = 2\Delta M_{1} \; ,
$$
$$
{ F^{0} - F^{2} - B^{0}   \over 2 I} = 2\Delta M_{2} \; ,
$$
$$
{ F^{0} - F^{2} - C^{0}   \over 2 I} = 2\Delta M_{3} \; ,
$$
$$
{ F^{2} - F^{3} \   -  C^{2}  + B^{3} \over 2I}  = 2 \Delta N^{1} \;,
$$
$$
 {F^{3} - F^{1} - A^{3} +  C^{1}  \over 2I}= 2 \Delta N^{2} \; ,
 $$
 $$
 {  F^{1}  - F^{2} - B^{1}  + A^{2} \over 2I} = 2 \Delta N^{3} \; .
 \eqno(28)
  $$

The answer can be presented in a concise form, if you go to a 3-dimensional vector  [10]
to parameterize the 4-vector Lorentz transformations:
$$
i{\bf q} = { {\bf k} \over  k_{0} } = {{\bf M} - i {\bf N} \over
\Delta },
\eqno(29)
$$

\noindent that is
$$
iq_{1} = 2{  (F^{0}- A^{0}) - F^{1}  -i [ ( F^{2} - F^{3}) -  (C^{2} - B^{3} )]
\over F^{0}    + (A^{1} - F^{1})  + (B^{2}- F^{2})  + (
C^{3}  - F^{3} )} \; ,
$$
$$
iq_{2} = 2{ (F^{0} - B^{0}) - F^{2}  - i [( F^{3} - F^{1} - (A^{3} - C^{1})]
 \over  F^{0}    + (A^{1} - F^{1})  + (B^{2}- F^{2})  + (
C^{3}  - F^{3} )} \; ,
$$
$$
iq_{2} = 2{ (F^{0} - C^{0} ) - F^{2}   - i [( F^{1} - F^{2}) - (B^{1} - A^{2} )]
\over  F^{0}    + (A^{1} - F^{1})  + (B^{2}- F^{2})  + (
C^{3}  - F^{3} ) } \; .
\eqno(30)
$$

To verify  these formulas let us consider a simple example. Let
a Mueller matrix of the Lorentz-type be
$$
M = L = \left | \begin{array}{cccc}
\cosh \beta  & 0 & 0 & \sinh \beta
  \\
0 & 1 & 0 & 0 \\
0 & 0 & 1 & 0  \\
\sinh \beta  & 0 & 0 & \cosh \beta \end{array} \right |,
$$

\noindent one can compute
$$
S^{0'}_{(0)} = F^{0}=I\, \cosh \beta\;, \; S^{1'}_{(0)} = F^{1}=0\;,
\; S^{2'}_{(0)} = F^{2}=0\;, \;  S^{3'}_{(0)} = F^{3}= I\,\sinh \beta\;,
$$
$$
 S^{0'}_{(1)} = A^{0}=I\, \cosh \beta\;, \; S^{1'}_{(1)} = A^{1}=I, \;
   S^{2'}_{(1)} = A^{2}=0, \;  S^{3'}_{(1)} = A^{3}=I\, \sinh \beta\;,
 $$
 $$
S^{0'}_{(2)} = B^{0}=I\, \cosh \beta\;,\;
 S^{1'}_{(2)} = B^{1}=0,\; S^{2'}_{(2)} = B^{2}=I,\; S^{3'}_{(2)} = B^{3}=I\, \sinh \beta\;,
$$
$$
S^{0'}_{(3)} = C^{0}=I\,(\cosh \beta+ \sinh \beta)\,,\qquad
S^{1'}_{(3)} = C^{1}=0\,,
$$ $$
S^{2'}_{(3)} =
C^{2}=0\,,\qquad S^{3'}_{(3)} = C^{3}=I\,(\cosh \beta+ \sinh
\beta)\,.
$$

Subsrtititu=ing the m into (30) we get an expected result
$$
q_{1}=0\,\qquad q_{2}=0\,,\qquad  q_{3}=-{\sinh \beta\over \cosh
\beta +1} =  i \mbox{tanh} {\beta \over 2} \; .
$$

\vspace{5mm}

Let us summarize the main result:  with assumption that an optical element is described by a Mueller
matrix of the Lorentzian type, a method to find a 3-dimensional
complex vector-parameter  for
 a corresponding Mueller matrix from results of
  four specially chosen polarization measurements has been elaborated.

\vspace{2mm}

Author is grateful   to V.M. Red'kov for moral support and advices.


\begin{thebibliography}{99}

\bibitem{Stokes-1852}
G.G. Stokes. On the composition and resolution of streams of polarized light from different sources.
Trans. Cambridge Phil. Soc. {\bf 9}, 399--419 (1852).

\bibitem{Mueller-1943}
H. Mueller. Memorandum on the polarization optics of the photo-elestic shutter.
Reporn No 2 of the OSRD project OEMsr576, Nov. 15 (1943).

\bibitem{Jones-1941}
R.C. Jones. New calculus for the treatment of optical systems. I.
 Description and discussion of the calculus. J. Opt. Soc. Amer. {\bf 31}, 488--493 (1941).

\bibitem{Hurwitz-1941}
H. Hurwitz, R.C. Jones. A new calculus for the treatment of optical systems. II.
Proof of three general equivalence theorems. J. Opt. Soc. Amer. {\bf 31}, 494--499 (1941).

\bibitem{Jones-1941(2)}
R.C. Jones. A new calculus for the treatment of optical systems. III.
The Sohncke Theory of optical activity. J. Opt. Soc. {\bf 31}, 500--503 (1941).

\bibitem{Jones-1947}
R.C. Jones. A new calculus for the treatment of optical systems. IV. Experimental determination of the matrix. J. Opt. Soc. {\bf 37}, 110--112 (1947).

\bibitem{Gel'fand-1963}
I.M. Gel'fand, R.A. Minlos, Z.Ya. Shapiro.
Representations of the rotation and
Lorentz groups and their applications. Pergamon, New York, 1963.


\bibitem{Fedorov-1958}
F.I. Fedorov.
Optics of anisotropic medias. Minsk, 1958.

\bibitem{Fedorov-1976}
F.I. Fedorov. The theory of hyrotropy. Minsk, 1976.

\bibitem{Fedorov-1979}
F.I. Fedorov,   The Lorentz group.  Moscow,  1979.

\bibitem{Berezin-1989}
A.V. Berezin,   Yu.A. Kurochkin,   E.A. Tolkachev.
Quternions  in relativistic physics.  Minsk,  1989.


\bibitem{Bogush-Red'kov-2008}
 A.A. Bogush, V.M. Red'kov. On Unique Parametrization of the Linear Group
 $GL(4.C)$ and Its Subgroups by Using the Dirac Algebra Basis. NPCS.  {\bf 11}, no 1,  1--24 (2008).

\bibitem{Bogush-2007}
A.A. Bogush, V.A. Dlugunovich, S.Ya. Zhukovich, Yu.A. Kurochin, V.N. Snopko.
Biquaternious and Mueller matrices.
Doklady of the National Academy of Sciences of Belarus.
{\bf  51}, no 5,   71--76 (2007).

\bibitem{Bogush-2008}
A.A. Bogush.
Mueller matrices in polarization optics.
Proc. of the Natl. Academy of Sciences of Belarus, Ser. Phys.-Math. Sci. {\bf  2}, 96--102 (2008).

\bibitem{Red'kov-arxiv/0906.2482}
V.M. Red'kov. Maxwell Equations in Media, Group Theory and Polarization of the Light.
 73 pages, arxiv/0906.2482.

\bibitem{Red'kov-2011}
V.M. Red'kov. Lorentz group and polarization of the light. Advances
in Applied Clifford Algebras.  {\bf 21},  203--220 (2011).

 \bibitem{Red'kov-2012}
 V.M. Red'kov, E.M. Ovsiyuk. Transitivity in the theory of the Lorentz group
 and the Stokes-Mueller formalism in optics. Reports  of Brest University. Series 4.
  Physics, mathematics. no 1,  (2012).

\bibitem{Snopko-1992}
V.N. Snopko.
Polarization characteristics  of optical radiation and methods of their measurement.
Minsk, 1992



\bibitem{Red'kov-bookI}
 V.M. Red'kov. The fields of the particles in a Riemannian space and the Lorentz group. Minsk, 2009.

\end{thebibliography}
\end{document}